\documentclass[twocolumn,showpacs,preprintnumbers,amsmath,amssymb]{revtex4}
\usepackage{graphicx}
\usepackage{dcolumn}
\usepackage{bm}
\def\pmb#1{\setbox0=\hbox{$#1$}%
  \kern-.025em\copy0\kern-\wd0
  \kern.05em\copy0\kern-\wd0
  \kern-.025em\raise.0433em\box0}
\def\parb{\pmb{\partial}}
\def\alt{\mathrel{\hbox{\rlap{\hbox{\lower4pt\hbox{$\sim$}}}\hbox{$<$}}}}
\begin{document}

\title{Numerical simulations of generic singularities}

\author{David Garfinkle}
\email{david@physics.uoguelph.ca}
\affiliation{Department of Physics, University of Guelph, Guelph, Ontario,
Canada N1G 2W1
\\ and Perimeter Institute for Theoretical Physics, 
35 King Street North, Waterloo Ontario, Canada N2J 2W9}

\begin{abstract}

Numerical simulations of the approach to the singularity in vacuum
spacetimes are presented here.  The spacetimes examined 
have no symmetries and can be regarded as representing the 
general behavior of singularities.  It is found that the singularity
is spacelike and that as it is approached, the spacetime dynamics 
becomes local and oscillatory.

\end{abstract}
\maketitle

A longstanding problem in general relativity has been to find the general
behavior of singularities.  Several results, both 
analytical\cite{alanreview}
and numerical\cite{bevreview}
have been obtained.  However, for most of the results,
the spacetimes have one or more symmetries.  Only for
scalar field matter have results been found when
there are no symmetries.\cite{larsandalan,harmonic}  For vacuum, 
Belinski, Lifschitz and 
Khalatnikov\cite{bkl} (BKL) 
have conjectured that the generic singularity is local,
spacelike and oscillatory.  This conjecture has been reformulated and
put more precisely by Uggla {\it et al}.\cite{jw} 

In this paper are presented numerical simulations of the approach to 
the singularity in vacuum spacetimes with no symmetries.  The results
support the BKL conjecture. 

The system evolved here is essentially that of reference\cite{jw} but
specialized to the vacuum case and 
with a slightly different choice of gauge.  Here the spacetime
is described in terms of a coordinate system ($t,{x^i}$) and a tetrad 
(${{\bf e}_0},{{\bf e}_\alpha}$)  where both the spatial coordinate 
index $i$ and 
the spatial tetrad index $\alpha $ go from 1 to 3.  Choose
${\bf e}_0$ to be hypersurface orthogonal with the relation between
tetrad and coordinates of the form
${{\bf e}_0} = {N^{-1}}{\partial _t}$ and  
${{\bf e}_\alpha} =
{{e_\alpha }^i}{\partial _i}$ 
where $N$ is the lapse and the shift is chosen to be zero.
Choose the spatial frame $\{ {{\bf e}_\alpha} \}$ to be
Fermi propagated along the integral curves of ${\bf e}_0$.
The commutators of the tetrad components are decomposed as follows:
\begin{eqnarray}
[{{\bf e}_0},{{\bf e}_\alpha}] &=& {{\dot u}_\alpha}{{\bf e}_0}
-(H {{\delta _\alpha}^\beta}
+{{\sigma _\alpha}^\beta})
{{\bf e}_\beta} 
\\
\left [ {{\bf e}_\alpha },{{\bf e}_\beta} \right ]  &=&
(2 {a_{[\alpha}}{{\delta _{\beta ]}}^\gamma}
+ {\epsilon _{\alpha \beta \delta }}{n^{\delta \gamma}}){{\bf e}_\gamma}
\end{eqnarray}
where $n^{\alpha \beta}$ is symmetric, and $\sigma ^{\alpha \beta}$ is 
symmetric and trace free. 

Scale invariant variables are defined as follows: 
$\{ {\parb_0},{\parb_\alpha} \} \equiv \{ 
{{\bf e}_0},{{\bf e}_\alpha} \} /H$ 
\begin{equation}
\{ {{E_\alpha}^i}, {\Sigma _{\alpha \beta }}, {A^\alpha} , 
{N_{\alpha \beta }} \} \equiv \{ {{e_\alpha}^i} , 
{\sigma _{\alpha \beta }} , {a^\alpha}, {n_{\alpha \beta}} \} /H
\end{equation} 
$q+1 \equiv - {\parb _0} \ln H $ and
${r_\alpha} \equiv - {\parb _\alpha} \ln H$.

Finally choose the lapse to be $N={H^{-1}}$.  The relation between
scale invariant frame derivatives and coordinate derivatives is
${\parb _0} ={\partial _t}$ and 
${\parb _\alpha} = {{E_\alpha }^i} {\partial _i}$.
From the vacuum Einstein equations one obtains the following 
evolution equations:
\begin{eqnarray}
{\partial _t} {{E_\alpha}^i} &=& {{F_\alpha}^\beta}{{E_\beta}^i}
\label{ev1}
\\
{\partial _t} {r_\alpha} &=& {{F_\alpha}^\beta}{r_\beta}+{\parb _\alpha}q
\label{ev2}\\
{\partial _t} {A^\alpha} &=& {{F^\alpha}_\beta}{A^\beta}+ 
{\textstyle \frac 1 2}{\parb  _\beta}{\Sigma ^{\alpha \beta}}
\label{ev3}
\\
\nonumber
{\partial _t} {\Sigma ^{\alpha \beta}} &=& (q-2) {\Sigma ^{\alpha \beta}}
- 2 {{N^{<\alpha}}_\gamma}{N^{\beta > \gamma}} + {{N_\gamma }^\gamma}  
{N^{<\alpha \beta >}} 
\\
\nonumber
&+& {\parb ^{<\alpha}}{r^{\beta >}} 
- {\parb ^{<\alpha}}{A^{\beta >}} 
+ 2{r^{<\alpha }}{A^{\beta >}} 
\\
&+& {\epsilon ^{\gamma \delta < \alpha}}({\parb _\gamma } - 2 {A_\gamma})
{{N^{\beta > }}_\delta}   
\label{ev4}
\\
{\partial _t}{N^{\alpha \beta}} &=& q {N^{\alpha \beta }} + 
2 {{\Sigma ^{(\alpha }}_\delta}{N^{\beta ) \delta }} - 
{\epsilon ^{\gamma \delta (\alpha }}{\parb _\gamma } 
{{\Sigma ^{\beta )}}_\delta}
\label{ev5}
\\
\nonumber
{\partial _t} q &=& \left [ 2 (q-2) + {\textstyle \frac 1 3} 
\left ( 2 {A^\alpha } - {r^\alpha}\right ) {\parb _\alpha}
- {\textstyle \frac 1 3} {\parb ^\alpha}{\parb _\alpha}
\right ] q 
\\
\nonumber
&-& {\textstyle \frac 4 3} {\parb _\alpha}{r^\alpha} + 
{\textstyle \frac 8 3}{A^\alpha}{r_\alpha} + {\textstyle \frac 2 3}
{r_\beta}{\parb _\alpha}{\Sigma ^{\alpha \beta}} 
\\
&-& 2 {\Sigma ^{\alpha \beta}}{W_{\alpha \beta}} 
\label{ev6}
\end{eqnarray}
Here angle brackets denote the symmetric trace-free part, and 
$F_{\alpha \beta }$ and $W_{\alpha \beta} $ are given by
\begin{eqnarray}
{F_{\alpha \beta }} &\equiv & q {\delta _{\alpha \beta}} - {\Sigma _{\alpha \beta}}
\\
\nonumber
{W_{\alpha \beta }} &\equiv & {\textstyle \frac 2 3}{N_{\alpha \gamma}}
{{N_\beta}^\gamma} 
- {\textstyle \frac 1 3} {{N^\gamma }_\gamma}
{N_{\alpha \beta }}
+ {\textstyle \frac 1 3} {\parb _\alpha} 
{A_\beta}
\\
&-& {\textstyle \frac 2 3} {\parb _\alpha} {r_\beta}
- {\textstyle \frac 1 3} 
{{\epsilon ^{\gamma \delta }} _\alpha } \left ( {\parb _\gamma}
- 2 {A_\gamma}\right ) {N_{\beta \delta}}  
\end{eqnarray}
In addition to the evolution equations, the variables satisfy constraint 
equations as follows:
\begin{eqnarray}
\nonumber
0 &=& {{({{\cal C}_{\rm com}})}^i _{\alpha \beta}} \equiv 
2 ( {\parb _{[\alpha }} - {r_{[\alpha}}-{A_{[\alpha}}){{E_{\beta ]}}^i}
\\
&-& {\epsilon _{\alpha \beta \delta}}{N^{\delta \gamma}}{{E_\gamma }^i}
\label{cn1}
\\
\nonumber
0 &=& {{\cal C}_{\rm G}} \equiv 1 + {\textstyle \frac 1 3} 
(2 {\parb _\alpha} - 2 {r_\alpha} - 3 {A_\alpha}){A^\alpha} -
{\textstyle \frac 1 6}{N_{\alpha \beta}}{N^{\alpha \beta}}
\\
&+&{\textstyle \frac 1 {12}} {{({{N^\alpha}_\alpha})}^2} -
{\textstyle \frac 1 6} {\Sigma _{\alpha \beta}}{\Sigma ^{\alpha \beta}} 
\label{cn2}
\\
\nonumber
0 &=& {{({{\cal C}_{\rm C}})}^\alpha} \equiv {\parb _\beta} 
{\Sigma ^{\alpha \beta}}+ 2 {r^\alpha} - {{\Sigma ^\alpha}_\beta}
{r^\beta} - 3 {A_\beta}{\Sigma ^{\alpha \beta}}
\\
&-&{\epsilon ^{\alpha \beta \gamma}}{N_{\beta \delta}}
{{\Sigma _\gamma}^\delta}   
\label{cn3}
\\
0 &=& {{\cal C}_q} \equiv q - {\textstyle \frac 1 3} {\Sigma ^{\alpha \beta}}
{\Sigma _{\alpha \beta}}+{\textstyle \frac 1 3} {\parb _\alpha }
{r^\alpha} - {\textstyle \frac 2 3}{A_\alpha}{r^\alpha}
\label{cn4}
\\
\nonumber
0 &=& {{({{\cal C}_{\rm J}})}^\alpha} \equiv ({\parb _\beta} - {r_\beta})
({N^{\alpha \beta}} + {\epsilon ^{\alpha \beta \gamma }}{A_\gamma})
\\
&-& 2 {A_\beta}{N^{\alpha \beta}}
\label{cn5}
\\
0 &=& {{({{\cal C}_{\rm W}})}^\alpha} \equiv [ {\epsilon ^{\alpha \beta 
\gamma }}({\parb _\beta} - {A_\beta}) - {N^{\alpha \gamma}}]{r_\gamma}
\label{cn6}
\end{eqnarray} 
We want a class of initial data satisfying these constraints that 
is general enough for our purposes
but simple enough to find numerically.  We use the York method\cite{jimmy} 
and choose
$H$ to be constant, $r^\alpha$ and $N_{\alpha \beta }$ to vanish and the
following form for the other variables: 
${{E_\alpha}^i}=({\psi ^{-2}}/H) {{\delta _\alpha}^i}, 
{A_\alpha} = - 2 {\psi ^{-1}}{\parb _\alpha} \psi $
and ${\Sigma _{\alpha \beta}} = ( - {\psi ^{-6}}/H) {\rm diag} (
{{\bar \Sigma}_1}, {{\bar \Sigma}_2}, {{\bar \Sigma}_3})$.
Then the constraints are satisfied provided ${\partial ^i}{{\bar \Sigma}_i}=0$
and $q={\frac 1 3} {\psi ^{-12}} {H^{-2}}
{{\bar \Sigma}^i}{{\bar \Sigma}_i}$
and 
\begin{equation}
{\partial ^i} {\partial _i} \psi = {\textstyle \frac 1 8}  (
6 {H^2} {\psi ^5} - {{\bar \Sigma}^i}{{\bar \Sigma}_i}{\psi ^{-7}} )     
\label{elliptic}
\end{equation}
We use the following solution for ${\bar \Sigma}_i$
\begin{eqnarray}
\nonumber
{{\bar \Sigma}_1}&=&{a_2}\cos y + {a_3} \cos z +{b_2}+{b_3}
\\
\nonumber
{{\bar \Sigma}_2}&=&{a_1} \cos x - {a_3} \cos z +{b_1}-{b_3}
\\
{{\bar \Sigma}_3} &=& -{a_1} \cos x - {a_2} \cos y - {b_1} - {b_2}
\end{eqnarray} 
where the $a_i$ and $b_i$ are constants.
We consider spacetimes
with topology ${T^3}\times R$ with each spatial slice having topology
$T^3$.  In terms of the coordinates we have $0\le x \le 2 \pi$ with
$0$ and $2\pi$ identified (and correspondingly for $y$ and $z$).

The numerical method used is as follows: each spatial direction
corresponds to $n+2$ grid points with spacing $dx=2\pi /n$.  The
variables on grid points $2$ to $n+1$ are evolved using the evolution
equations, while at points $1$ and $n+2$ periodic boundary conditions
are imposed.  The initial data is determined once equation 
(\ref{elliptic}) is solved.  This is done using the conjugate gradient
method.\cite{Saul}  The evolution proceeds using equations 
(\ref{ev1}-\ref{ev6}) with the exception that the term 
$(5-2q) {{\cal C}_q}$ is
added to the right hand side of equation (\ref{ev6}) to prevent the
growth of constraint violating modes.  Spatial derivatives are 
evaluated using centered differences, and the evolution is done using
a three step iterated Crank-Nicholson method\cite{ICN} (a type of 
predictor-corrector method).  
In equation (\ref{ev6}) the highest spatial derivative term 
is $-{\frac 1 3} {\parb ^\alpha}{\parb _\alpha}q$ which
gives this equation the form 
of a diffusion equation.  Note that diffusion equations can only be
evolved in one direction in time, in this case the negative
direction which corresponds to the approach to the singularity.
Stability of numerical evolution of diffusion equations generally
requires a time step proportional to the sqare of the spatial step.  
However, the constant of proportionality depends on the coefficient
of the second spatial derivative.  To ensure stability, we define
${E_{\rm max}}$ to be the maximum value of $|{{E_\alpha}^i}|$ (over all
space and over all $\alpha $ and $i$) and then define 
$d{t_1} \equiv - {\frac 1 4} {{(dx/{E_{\rm max}})}^2}$ and 
$d{t_2} \equiv - {\frac 1 8} dx$.  The time step $dt$ is then chosen
to be whichever of $d{t_1}$ and $d{t_2}$ has the smaller magnitude. 

Before presenting numerical results, it is helpful to consider what
behavior to expect as the singularity is approached (that is 
as $t\to -\infty$).  First denote the eigenvalues of 
${\Sigma ^\alpha}_\beta$ by $({\Sigma_1},{\Sigma_2},{\Sigma_3}).$
Then suppose that at sufficiently early
times the time averages of $q-{\Sigma_i}$ are all positive.  
Then the time averages of the eigenvalues of 
${F^\alpha}_\beta$ are all positive.
Since we are evolving in the negative time direction, this should lead
(through equation (\ref{ev1}))
to an exponential decrease in ${E_\alpha}^i$.  However, since all
spatial derivatives appear in the equations through 
${\parb _\alpha} = {{E_\alpha }^i} {\partial _i}$ we would 
expect the spatial derivatives to become negligible.  That is, at
each spatial point the dynamics becomes that of a spatially homogeneous
cosmology: the approach to the singularity is local.  
Note that this does not mean that the spacetime is becoming homogeneous.
Spatial variation is not becoming small; however since all spatial 
derivatives appear in the evolution equations through
${{E_\alpha }^i}{\partial _i}$ and since ${E_\alpha}^i$ is becoming
small, the effect of the spatial derivatives on the evolution is 
becoming negligible. 
the positivity of the time averages of the eigenvalues of 
${F^\alpha}_\beta$ should also lead 
(through equations (\ref{ev2}-\ref{ev3})) to exponential decrease
in $r_\alpha$ and $A^\alpha$.  Thus we expect that as the singularity
is approached, the dynamics is described by a much simpler version of
evolution equations (\ref{ev1}-\ref{ev6}) and constraint equations
(\ref{cn1}-\ref{cn6}) where $r_\alpha$ and $A^\alpha$ and all 
spatial derivatives are dropped.   

This simpler set of equations describes homogeneous cosmologies,
and has been treated extensively in the literature on such 
spacetimes.\cite{bkl,we} Here we simply summarize the important
attributes. 
First, from equation (\ref{cn3}) it follows that 
${\Sigma ^\alpha}_\beta$ and ${N^\alpha}_\beta$ commute and
therefore have a common basis of eigenvectors.  From equations
(\ref{ev4}-\ref{ev5}) it follows that this basis (which we will
call the asymptotic frame) is not changed 
under time evolution.  One can therefore express equations 
(\ref{ev4}-\ref{ev5}) in the asymptotic frame yielding equations 
for the ${\Sigma _i}$ (eigenvalues of ${\Sigma ^\alpha}_\beta$)
and the $N_i$ (eigenvalues of ${N^\alpha}_\beta$).   During a time
period where all the $N_i$ are negligible, it follows that all the
$\Sigma_i$ are constant.  Such a time period is called a Kasner epoch.
During a Kasner epoch, two of the $N_i$ are decaying and one is growing.
The growing eigenvalue of ${N^\alpha}_\beta$ eventually gives rise
to a transition (called a ``bounce'') to another Kasner epoch.  It is
helpful to define a quantity $u$ by
\begin{equation}
{{\Sigma^\alpha}_\beta}{{\Sigma^\beta}_\gamma}{{\Sigma^\gamma}_\alpha}
= 6 - {\frac {81 {u^2} {{(1+u)}^2}} {{(1+u+{u^2})}^3}} 
\end{equation}    
(there is a unique $u \ge 1$ provided the quantity on the left hand side of 
the equation is between -6 and 6).  
The quantity $u$ is constant in each Kasner epoch and changes from epoch
to epoch as follows: $u \to u-1$ if $u \ge 2$ and $u \to 1/(u-1)$ if
$1<u\le 2$.  This rule is called the $u$ map. 
\begin{figure}
\includegraphics[scale=0.6]{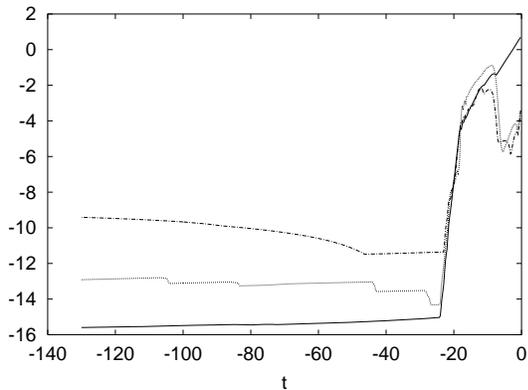}
\caption{\label{fig1} maximum values of 
$\ln |{{E_\alpha }^i}|$ (solid line), $\ln |{r_\alpha}|$ (dotted line) and 
$\ln |{A^\alpha}|$ (dot-dashed line)  {\it vs} time}
\end{figure}

\begin{figure}
\includegraphics[scale=0.6]{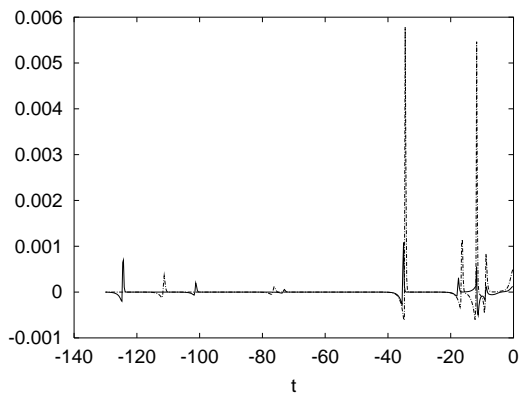}
\caption{\label{fig2}Constraint ${\cal C}_q$ {\it vs} time, for 
$n=50$ (solid line) and $n=25$ (dot-dashed line)}
\end{figure}

We now turn to the numerical simulations.  All runs were done in
double precision on a SunBlade
2000 with $n=50$ (except for an examination of resolution
which also used $n=25$).
The equations were evolved from $t=0$ to $t=-130$.
The initial value of $H$ was $\frac 1 3$ corresponding to an initial trace
of extrinsic curvature equal to $-1$.  The $b_i$ were chosen with ${b_3}=0$
and $b_1$ and $b_2$ given so that the spacetime would be a Kasner spacetime
with $u=2.3$ if the $a_i$ vanished.  For the runs presented here the
$a_i$ were given as $(0.2,0.1,0.04)$.  

We would like to know whether ${{E_\alpha}^i}, {r_\alpha}$
and $A^\alpha$ become negligible as the singularity is approached.  
In figure \ref{fig1} are
plotted the maximum values (over all space, $\alpha$ and $i$) of  
$\ln |{{E_\alpha}^i}|, \ln |{r_\alpha}|$ and $\ln |{A^\alpha}|$ as 
functions
of time.  Note the steep decrease in these quantities near $ t \sim -20$.
This indicates that 
$r_\alpha , \; {A_\alpha} $ 
and the spatial 
derivatives become negligible for $t \alt -20$.
(the failure of the quantities plotted in figure 1 to continue to 
decrease 
is most likely due to unresolved small scale spatial structure to be 
discussed below).
Thus the interesting part of the dynamics can be seen
by looking at the development of the variables at a single point as a
function of time.  We now present data of that form. 
The behavior at
the spatial point
chosen is typical.    

The behavior of a constraint 
is presented in figure \ref{fig2}.  Here what is plotted is 
$\ln |{{\cal C}_q} |$ 
at the spatial point as a function of 
time. The solid line is for the $n=50$ run and the dot-dashed line is 
for the $n=25$ run. 
The finer resolution yields a smaller value for the constraint; 
but the resolution is not good enough to be in the convergent regime.
Note that the time
dependence of the two runs becomes increasingly out of sync.  This is
due to the chaotic nature of homogenous cosmologies.\cite{bevreview}
Also note that the shape of the constraint for the two runs does not
completely match.  This may be due to unresolved small scale structure to
be discussed below.
Similar results were obtained for the other constraints.

\begin{figure}
\includegraphics[scale=0.6]{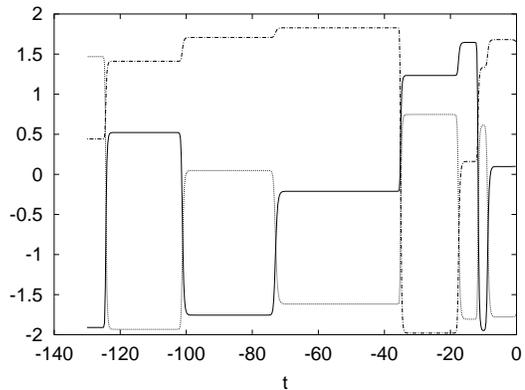}
\caption{\label{fig3} components of $\Sigma_{\alpha \beta}$ 
{\it vs} time, in the 
asymptotic frame: $\Sigma _1$ (solid line), $\Sigma _2$ (dotted line)
and $\Sigma_3$ (dot-dashed line)}
\end{figure} 

\begin{figure}
\includegraphics[scale=0.6]{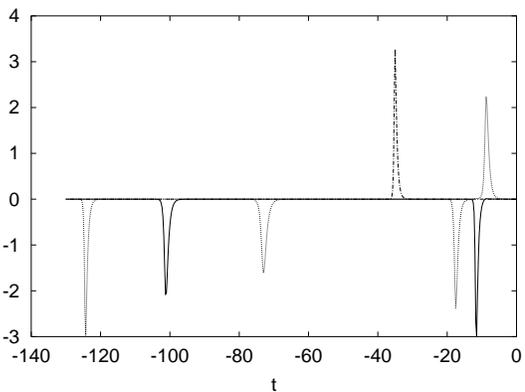}
\caption{\label{fig4} components of $N_{\alpha \beta}$ {\it vs} time, in the
asymptotic frame: $N_1$ (solid line), $N_2$ (dotted line) and $N_3$
(dot-dashed line)}
\end{figure}

\begin{figure}
\includegraphics[scale=0.6]{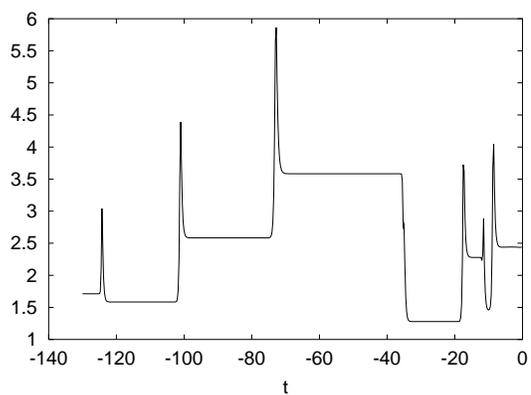}
\caption{\label{fig5} $u$ {\it vs} time}
\end{figure}

Figures \ref{fig3} and \ref{fig4} show respectively 
the diagonal components of $\Sigma _{\alpha \beta}$ and 
$N_{\alpha \beta}$ in the asymptotic frame.  Though not shown here, I 
have also examined the off-diagonal components of both 
$\Sigma _{\alpha \beta}$ and $N_{\alpha \beta}$ in this frame.  For times 
$t \alt -20$ these off-diagonal components become negligible.  This 
demonstrates that for these times the asymptotic frame is essentially 
unchanged and the quantities in figures \ref{fig3} and 
\ref{fig4} are eigenvalues of
${\Sigma ^\alpha}_\beta$ and ${N^\alpha}_\beta$ respectively.
Note that for $t\alt -20$ the behavior of the components of 
$\Sigma _{\alpha \beta}$ consists of epochs where they are constant
punctuated by short bounces where they change rapidly.  Note too that
for $t \alt -20$ the behavior of the components of $N_{\alpha \beta}$
is that they are negligible during the epochs of 
constant $\Sigma _{\alpha \beta}$ and that one component of $N_{\alpha \beta}$
becomes non-negligible at each bounce.  This is exactly what we would
expect from the approximation of using the equations for homogeneous 
spacetimes.   

We now turn to the behavior of the quantity $u$.  In figure \ref{fig5} is 
plotted $u$ as a function of time.  Note that $u$ undergoes a series of
bounces when 
the components of $\Sigma_{\alpha \beta}$ do.  
The sequence of values of $u$ begining at $t \sim -20$ is 
$ 1.279, 3.583, 2.584, 1.584, 1.712$.  This sequence obeys the $u$ map.

In summary, these simulations provide strong support for the BKL conjecture.
The initial data evolved have no symmetry and can be regarded as generic.
The evolution shows that spatial derivatives become negligible and the
time dependence goes over to the well studied behavior of a general
homogeneous cosmology.  This dynamics is oscillatory consisting of a series
of Kasner epochs punctuated by short bounces.  The details of the oscillations
given by the behavior of the quantity $u$ are in agreement with what would
be expected for locally homogeneous spacetimes.  

We now consider what remains to be done on this subject.  First recall 
that bounces occur when
one of the $N_i$ grows exponentially.  If that $N_i$ initially vanishes
on a surface $\cal S$ then we would expect bounces on either side of 
$\cal S$, but not on $\cal S$ itself.  This gives rise to a small scale
structure which can be seen (crudely) in the results of these simulations,
and which has been well studied in the 
Gowdy spacetimes.\cite{meandbeverly,alanandmarsha} 
A corresponding study for the case of no symmetry 
will require high resolution and is work in progress.

The present work only treats vacuum
spacetimes.  There is a general expectation that for most types of
matter (a scalar field is an exception) the influence of the matter
on the dynamics should become negligible as the singularity is
approached.  The formalism of reference\cite{jw} is for a fluid with
equation of state $P=k\rho$ for constant $k$.  Thus a fairly straightforward
generalization of the simulations reported here would be to do simulations
of the approach to the singularity for $P=k\rho$ perfect fluid.  

Another question is whether there are any residual effects of
the spatial derivatives
as the singularity is approached.  
Comparison of the full evolution to one where the spatial 
derivatives are set to zero after a time $t_0$ yields differences:
the sequence of bounces is the same, but they occur more rapidly
in the full evolution.  This may be due to the spatial derivatives'
increasing the values of the $N_i$ and thus hastening the time 
when the $N_i$ become large enough to cause a bounce.

Finally, note that this simulation is for a spatially closed universe.
Since the result is that the dynamics become local as the singularity
is approached, these results should describe at least a portion of 
the singularity
in any generic spacetime, including asymptotically flat spacetimes  
that undergo gravitational collapse to form black holes.  Nonetheless,
there is a body of work\cite{eric,eanna} 
that indicates that when a black hole forms, that
portion of the singularity that is near the event horizon is null
(or asymptotically null).  It would be good to extend the methods  
presented here so that they are able to treat collapse in asymptotically
flat spacetimes and the near horizon properties of the singularities
formed. 

I would like to thank Mark Miller, Beverly Berger, Woei-Chet Lim, 
John Wainwright, Lars Andersson, Jim Isenberg and G. Comer Duncan for helpful 
discussions.  This work was partially supported by NSF grant
PHY-0244683 to Oakland University.


\begin{thebibliography}{2}

\bibitem{alanreview}
for a review see A. Rendall, ``Theorems on existence and global
dynamics for the Einstein equations''
Living Reviews in Relativity (2002-6)

\bibitem{bevreview}
for a review see B. Berger, ``Numerical Approaches to spacetime singularities''
Living Reviews in Relativity (2002-1)

\bibitem{larsandalan}
L. Andersson and A. Rendall, Commun. Math. Phys. {\bf 218}, 479 (2001)

\bibitem{harmonic}
D. Garfinkle, Phys. Rev. {\bf D65}, 044029 (2002)

\bibitem{bkl}
V. Belinskii, I. Khalatnikov and E. Lifschitz, Sov. Phys. Usp. {\bf 13},
745 (1971)

\bibitem{jw}
C. Uggla, H. van Elst, J. Wainwright and G.F.R. Ellis, Phys. Rev.
{\bf D68}, 103502 (2003)

\bibitem{jimmy}
J.W. York, Phys. Rev. Lett. {\bf 26}, 1656 (1971)

\bibitem{Saul}
W. Press, S. Teukolsky, W. Vetterling and B. Flannery, 
{\it Numerical Recipes in FORTRAN}, second edition 
(Cambridge University Press, Cambridge, 1992)

\bibitem{ICN}
M. Choptuik, in {\it Deterministic Chaos in General Relativity,} 
edited by D. Hobill, A. Burd and A. Coley (Plenum, New York, 1994),
pp. 155-175 

\bibitem{we}
``Dynamical systems in cosmology'' Edited by J. Wainwright and G.F.R. Ellis,  
Cambridge University Press, 1997

\bibitem{meandbeverly}
B. Berger and D. Garfinkle, Phys. Rev. {\bf D57}, 4767 (1998)   

\bibitem{alanandmarsha}
A. Rendall and M. Weaver, Class. Quantum Grav. {\bf 18}, 2959 (2001)

\bibitem{eric}
E. Poisson and W. Israel, Phys. Rev. {\bf D41}, 1796 (1990)

\bibitem{eanna}
A. Ori and E. Flanagan, Phys. Rev. {\bf D53}, 1754 (1996) 

\end{thebibliography}
\end{document}